\documentclass[prd,reprint,nofootinbib,amsmath,amssymb,aps,floatfix]{revtex4-1}
\usepackage{amssymb,amsmath,graphicx,color}
\usepackage[linktoc=page]{hyperref}
\usepackage{amsfonts}
\numberwithin{equation}{section}

\begin{document}
\title{Asymptotic Entropy Bounds}
\author{Raphael Bousso}%
 \email{bousso@lbl.gov}
\affiliation{Center for Theoretical Physics and Department of Physics\\
University of California, Berkeley, CA 94720, USA 
}%
\affiliation{Lawrence Berkeley National Laboratory, Berkeley, CA 94720, USA}
\begin{abstract}
  We show that known entropy bounds constrain the information carried off by radiation to null infinity. We consider distant, planar null hypersurfaces in asymptotically flat spacetime. Their focussing and area loss can be computed perturbatively on a Minkowski background, yielding entropy bounds in terms of the energy flux of the outgoing radiation. In the asymptotic limit, we obtain boundary versions of the Quantum Null Energy Condition, of the Generalized Second Law, and of the Quantum Bousso Bound. 
\end{abstract}
\maketitle
\tableofcontents

\section{Introduction}
\label{intro}

\subsection{Entropy Bounds}
Gravitational entropy bounds~\cite{Bek74,Bek81,Tho93,Sus95,FisSus98,CEB1,FMW,RMP,BouFla03,StrTho03,Cas08,Wal11,BouCas14a,BouCas14b,BouFis15a,BouFis15b} are of the general form 
\begin{equation}
S\leq \frac{\Delta A}{4 G\hbar}~,
\label{eq-bound}
\end{equation}
where $S$ is a suitable measure of the quantum information or entropy carried by matter systems, and $\Delta A$ is the area of a surface or the difference between two surface areas. $G$ and $\hbar$ are Newton's and Planck's constants; we set $c=k_B=1$.


The holographic scaling with area is surprising and conflicts with locality. However, there is considerable evidence that Eq.~(\ref{eq-bound}) holds in Nature, if $S$ is taken to be the entropy of matter systems crossing a nonexpanding null hypersurface called light-sheet~\cite{CEB1,FMW,BouFla03,RMP}. This is called the covariant entropy bound, or Bousso bound. In its most general form it remains a conjecture about the semi-classical regime. Its proof will likely require a full quantum theory of gravity. However, in the weak-gravity limit, it has already been possible to prove Eq.~(\ref{eq-bound}). 

As gravity becomes weak, $G\to 0$, one might expect entropy bounds to become trivial, since Newton's constant $G$ appears in the denominator in Eq.~(\ref{eq-bound}). Remarkably, this is not the case, if the light-rays are chosen to be parallel at $O(G^0)$. 
The area difference $\Delta A$ on null hypersurfaces (such as event horizons or light-sheets) then results entirely from the focussing of the light-rays by matter and radiation. Thus, $\Delta A$ will be proportional to $G$~\cite{Bou03}, and Newton's constant drops out of Eq.~(\ref{eq-bound}) as $G\to 0$. 

The bounds can then be expressed in terms of integrals over the energy flux ${\cal T}$ that causes focussing. ${\cal T}$ is the matter stress tensor component in the light-sheet direction, plus a shear-squared term that is associated with gravitational radiation. Using precise definitions of $S$~\cite{SusTho93,StrTho03,MarMin04,Cas08,Wal10,Wal11,BouCas14a,BouCas14b,BouFis15a}, the $G\to 0$ limit yields novel, highly nontrivial statements about quantum field theory: a Quantum Bousso Bound (QBB)~\cite{BouCas14a}, and the Quantum Null Energy Condition (QNEC)~\cite{BouFis15a}. One can also consider the Generalized Second Law in this limit~\cite{Wal10,Wal11}. In some cases, the weak gravity bounds can be proven rigorously within quantum field theory~\cite{Wal10,Wal11,BouCas14a,BouCas14b,BouFis15b,KoeLei15}.

In the present paper, we will explore a different limit, in which $G$ is held fixed but gravity nevertheless becomes weak: the limit of distant null planes in asymptotically flat space. We find that the above weak-gravity bounds apply in this setting. (In particular, this implies that all relevant quantities can be computed on a Minkowski background; we need not consider curved metrics explicitly.) Taking the limit as the null planes recede to future null infinity, we will show that each known weak-gravity entropy bound implies a corresponding bound on the information arriving at the conformal boundary, ${\cal I}^+$.  
\begin{figure}[ht]
\includegraphics[width=.4 \textwidth]{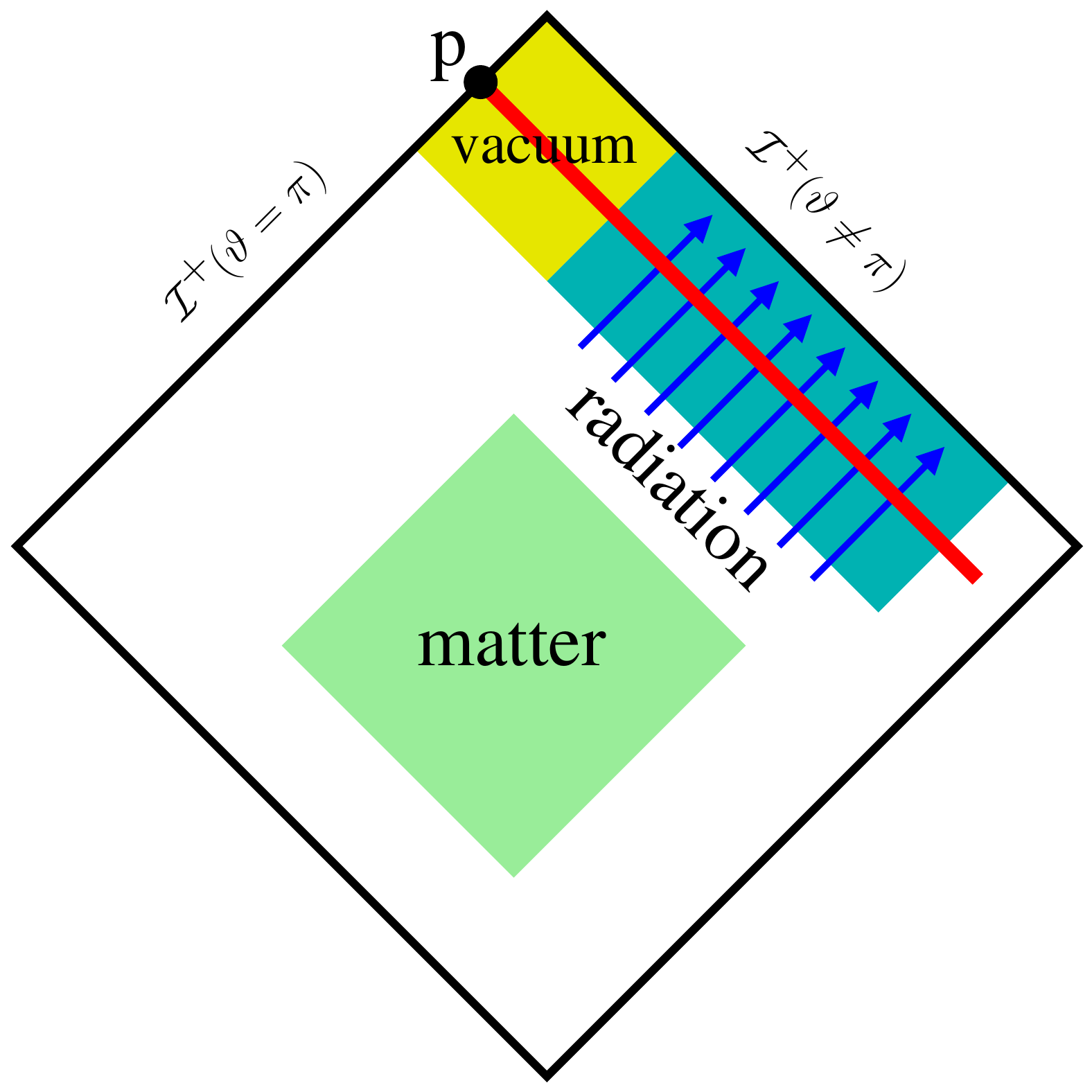}
\caption{Conformal diagram of an asymptotically flat spacetime. The left boundary represents only $\vartheta=\pi$; the right boundary represents all other angles on ${\cal I}^+$. The light-sheet $H(u_p)$ (red thick line) is the boundary of the past of a point $p$ at $(u_p, \vartheta=\pi)$ on ${\cal I}^+$. For large $u_p$, only outgoing radiation (blue arrows) passes through $H(u_p)$. All massive systems are assumed to decay into radiation in finite time, so the yellow region at the top is empty.}
\label{fig-zones}
\end{figure}

\subsection{Outline and Summary of Results}

We review relevant known entropy bounds in Sec.~\ref{sec-gravity}, and their standard weak gravity limits in Sec.~\ref{sec-weak}. Our main results appear in Sec.~\ref{sec-asymptotic} and Sec.~\ref{sec-scri}.

\begin{figure}[ht]
\includegraphics[width=.4 \textwidth]{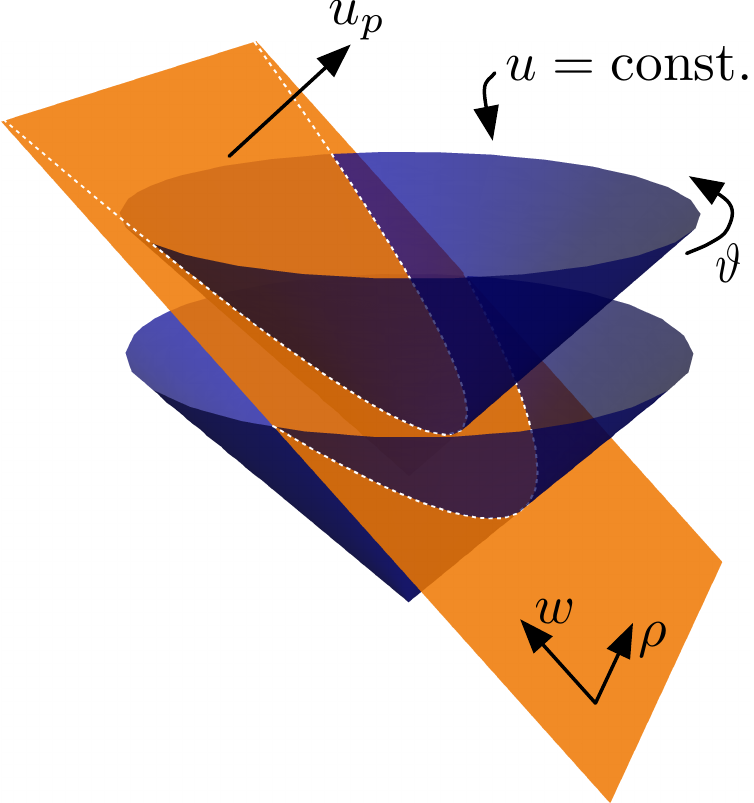}
\caption{The light-sheet $H(u_p)$ (orange) is approximately planar. The outgoing radiation is approximately radial (blue cones). Its focussing effect on $H(u_p)$ depends on the angle $\vartheta$ where it strikes $H(u_p)$, like $\cos^4(\vartheta/2)$. In the limit as $u_p\to\infty$, this factor cancels against the transformation between bulk and boundary affine parameters $u$ and $w$. The resulting entropy bounds on ${\cal I}^+$ do not depend on the orientation of $H(u_p)$. {\em Image credit: Z.~Fisher}}
\label{fig-conical}
\end{figure}

In Sec.~\ref{sec-asymptotic}, we consider a one-parameter family of light-sheets $H(u_p)$ in asymptotically flat spacetime (Figs.~\ref{fig-zones}, \ref{fig-conical}). The light-sheets are constructed so as to be approximately planar; at leading order they are given by the null planes $t-z=u_p$ in Minkowski space.  As $u_p$ becomes large, the light-sheets $H(u_p)$ approach future null infinity, ${\cal I}^+$. We show that focussing and area loss can be computed at order $G/u_p^2$, simply by applying the focussing equation to radiation propagating through $H(u_p)$ on the trivial $O((G/u_p^2)^0)$ background. 

We thus find that all weak-gravity entropy bounds apply directly to the light-sheets $H(u_p)$, for large $u_p$ but fixed $G$. This means that weak-gravity entropy bounds such as the QNEC and the QBB limit the flow of information out of arbitrarily large, isolated systems with arbitrary self-gravity, in terms of the energy flux density ${\cal T}$. 

In Sec.~\ref{sec-scri} we take the limit as $u_p\to\infty$, and we show that $H(u_p)$ becomes ${\cal I}^+$  in the ``unphysical spacetime'' obtained by conformal rescaling (the Penrose diagram). ${\cal I}^+$ is a light-sheet with vanishing expansion. This is perhaps counterintuitive: the $H(u_p)$ have planar cross-sections at $O(G^0)$, whereas the spatial cross-sections of ${\cal I}^+$ are conventionally chosen as spheres of unit radius. Yet, the limit can be established. In particular, the retarded time $u$, which is an affine parameter on ${\cal I}^+$, becomes an affine parameter on $H(u_p)$ in the limit.

We define a boundary energy flux $\hat {\cal T}$ by an angle-dependent rescaling of ${\cal T}$. For the matter part of $\hat {\cal T}$, we show explicitly that our definition reduces to the standard boundary ``matter stress tensor,'' which characterizes nongravitational radiation arriving on ${\cal I}^+$. This quantity is manifestly finite and independent of the orientation of the $H(u_p)$. It can be shown~\cite{BouKoe16} that the corresponding rescaling of the shear on $H(u_p)$ is equal to the Bondi news.

We identify a spatial cross-section, or ``cut,'' on every $H(u_p)$, corresponding to a given cut $\hat\sigma$ on ${\cal I}^+$, such that every cut partitions the outgoing radiation in the same way as $u_p\to\infty$. We can consider the entropy of one part of the radiation, i.e., the von Neumann entropy of the reduced quantum state obtained by restricting the global state to the portion of $H(u_p)$ on one side of the cut. By applying entropy bounds to each $H(u_p)$ and taking the limit, we obtain entropy bounds on ${\cal I}^+$ in terms of the finite boundary energy flux $\hat {\cal T}$. 

Our strongest result is the Boundary Quantum Null Energy Condition,
\begin{equation}
\frac{1}{\delta\Omega} \frac{d^2}{du^2} \hat S_\text{out}[\hat\sigma,\Omega] \leq \frac{2\pi}{\hbar}~ 
\hat{\cal T}~.
\label{eq-qnecscri0}
\end{equation}
We obtain a Boundary Generalized Second Law, both in differential form
\begin{equation}
-\frac{1}{\delta\Omega} \frac{d}{du}\hat S_\text{out}[\hat\sigma;\Omega] \leq 
\frac{2\pi}{\hbar}\int_{\hat\sigma}^\infty du~ \hat{\cal T}~,
\label{eq-gslscri0}
\end{equation}
and in integral form
\begin{equation}
\hat S_C[\hat\sigma_2]\leq
\frac{2\pi}{\hbar}\!\int_{\hat\sigma_2}^{\infty}\!\!\! d^2\Omega\, du\, 
[u-u_2(\Omega)]\, \hat{\cal T}~.
\label{eq-gslintscri0}
\end{equation}
(See the main text for detailed definitions. Divergences of the entanglement entropy cancel in the derivatives and subtractions.) Finally, we derive a Boundary Quantum Bousso Bound, which refers to the vacuum-subtracted entropy of a finite affine interval and involves additional subtleties.

\subsection{Related Work} Recently, Kapec, Raclariu, and Strominger (KRS) conjectured an asymptotic entropy bound on ${\cal I}^+$~\cite{KapRac16}. Null surfaces with approximately spherical cross-sections (past light-cones) are considered. In this setup, existing entropy bounds become trivial in the asymptotic limit, since areas and area differences diverge. Ref.~\cite{KapRac16} proposes an additional subtraction to cancel this divergence, which amounts to conjecturing a novel entropy bound. The definition of the entropy appearing in this bound was left to future work, so the conjecture is (for now) that an appropriate definition can be found~\cite{KapRac16}.

The present work takes a different approach: we consider planar light-sheets, on which area differences between cuts remain finite. This allows us to exploit standard bulk bounds on fairly rigorously defined measures of entropy, for which proofs or substantial evidence have already been found. A potential downside is that the planar light-sheets carry an orientation, so one might expect to obtain a separate boundary statement for each orientation angle. However, when we take the limit as $u_p\to\infty$ and express the bounds in terms of finite rescaled quantities on ${\cal I}^+$, we find that the results are independent of the angle chosen. Thus we obtain a unique boundary version of each type of bound. 

It is not possible to determine whether the KRS conjecture implies, or is implied by, any of the bounds derived here, because no definition of the entropy was given.
Formally, Eq. (66) of Ref.~\cite{KapRac16} can be compared to a special case of our results: the integrated form of the GSL, Eq.~(\ref{eq-gslintscri0}), with the further choice $\sigma_1\to\infty$. The right side of Eq.~(\ref{eq-gslintscri0}) then reduces (up to sign conventions) to the quantity denoted $A^\Sigma_F$ in~\cite{KapRac16}. The KRS conjecture contains an extra surface term of indefinite sign; see Eq. (22) in~\cite{KapRac16}. Perhaps it is possible to define the entropy in the KRS conjecture so that it differs from the l.h.s.\ of Eq.~(\ref{eq-gslintscri0}) by the same term; if so, the KRS conjecture would reduce to Eq.~(\ref{eq-gslintscri0}). This question will be considered in a separate publication~\cite{BouKoe16}, where we provide a more detailed treatment of the contributions from gravitons.

\section{Entropy Bounds With Gravity}
\label{sec-gravity}

In this section, we state the Generalized Second Law (Sec.~\ref{sec-gsl}) in a rigorous form, and we review the Quantum Focussing Conjecture Sec.~\ref{sec-qfc}. We begin by defining the generalized entropy and the quantum expansion, in Sec.~\ref{sec-defs}.

\subsection{Definitions}
\label{sec-defs}

Let $H$ be a null hypersurface with affine parameter $w$ and transverse coordinates $y$. Let $\sigma$ be a spatial cross-section of $H$, or {\em cut}. For example, $\sigma$ can be specified by a function $w(y)$.  The {\em generalized entropy}~\cite{Bek72,Bek73,Bek74} is the functional
\begin{equation}
S_\text{gen}[\sigma] \equiv S_\text{out}[\sigma]+\frac{A[\sigma]}{4G\hbar}+ \ldots~,
\label{eq-sgendef}
\end{equation}
where $S_\text{out}$ is the von Neumann entropy of the density operator of the quantum fields restricted to one side of the cut $\sigma$. (It is assumed here that $\sigma$ splits a Cauchy surface. It does not matter which side is chosen~\cite{BouFis15a}.)\footnote{For helpful figures illustrating the definitions in this section, see for example Ref.~\cite{BouFis15a}.}

Notably, $S_\text{gen}$ is better defined than either $S_\text{out}$ or $A[\sigma]/4G\hbar$ separately~\cite{SusTho93}. The leading divergence in the exterior entropy is proportional to $A[\sigma]$. The Bekenstein-Hawking term can be regarded as a counterterm. The ``$\ldots$'' in Eq.~(\ref{eq-sgendef}) stands for additional geometric counterterms, e.g., higher curvature terms, which cancel subleading divergences of the von Neumann entropy. If the exterior region consists of well-isolated systems far from $H$, the two terms on the right hand side become separately well-defined, with $S_\text{out}$ the standard thermodynamic entropy of the systems and $G$ the ``infrared'' value of Newton's constant. (See Ref.~\cite{BouFis15a} for a brief review and references.)
Consider a deformation of $\sigma$ by an infinitesimal distance $dw$ along a neighborhood of the generator $y$, of infinitesimal area ${\cal A}$. The change in $S_\text{gen}$ will be proportional to ${\cal A}$ and to $dw$. The {\em quantum expansion} of $\sigma$ at $y$ is defined~\cite{BouFis15a} as
\begin{equation}
\Theta[\sigma; y] = \frac{4G\hbar}{{\cal A}} S_\text{gen}'[\sigma;y]~,
\label{eq-qe}
\end{equation}
where the prime denotes $d/dw$. The limit as ${\cal A}\to 0$ is implicit wherever ${\cal A}$ appears. The quantum expansion depends both on the cut $\sigma$, and on where $\sigma$ is deformed (at $y$). 

Using Eq.~(\ref{eq-sgendef}), the quantum expansion can be expressed as
\begin{equation}
\Theta[\sigma; y] = \frac{4G\hbar}{{\cal A}}  S_\text{out}'[\sigma;y] + \theta[\sigma;y]~,
\end{equation}
where
\begin{equation}
\theta[\sigma;y] = \frac{{\cal A}'}{\cal A}
\label{eq-thetadef}
\end{equation}
is the classical expansion, i.e., the trace of the null extrinsic curvature of $\sigma$ in $H$ at $y$. (The definitions of ${\cal A}$ and the prime are given in the previous paragraph.) Unlike the quantum expansion, $\theta$ is local: it does not depend on the cut $\sigma$ away from $y$.

\subsection{Generalized Second Law}
\label{sec-gsl}

Now let us specialize to a null hypersurface $H$ that is a causal horizon (i.e., the boundary of the past of an inextendible timelike or null curve). The {\em Generalized Second Law} (GSL)~\cite{Bek72} on a future causal horizon is the conjecture that
\begin{equation}
\Theta[\sigma; y]\geq 0
\label{eq-gsl}
\end{equation}
for any future-directed deformation at $y$ of any cut $\sigma$ of $H$. (See Ref.~\cite{Wal13,BouFis15a} for the present formulation.) That is, the generalized entropy will not decrease towards the future. 

Eq.~(\ref{eq-gsl}) generalizes both the ordinary Second Law of thermodynamics (to the case where horizons are present), and Hawking's area theorem for event horizons (to the case where the Null Energy Condition need not hold). In cases where the generalized entropy can be separated into area and exterior entropy, Eq.~(\ref{eq-gsl}) becomes
\begin{equation}
-\frac{S_\text{out}'[\sigma;y]}{\cal A} \leq \frac{\theta[\sigma;y]}{4G\hbar}~.
\label{eq-gslsep}
\end{equation}

Consider two cuts of the horizon such that $\sigma_1$ is nowhere to the past of $\sigma_2$. Integration of Eq.~(\ref{eq-gsl}) gives the integral form of the GSL: 
\begin{equation}
S_\text{gen}[\sigma_1]-S_\text{gen}[\sigma_2]\geq 0~.
\label{eq-gslint}
\end{equation}

Specializing to the separable case, Eq.~(\ref{eq-gslint}) becomes
\begin{equation}
S_\text{out}[\sigma_2]-S_\text{out}[\sigma_1]\leq \frac{A[\sigma_1]-A[\sigma_2]}{4G\hbar} ~.
\label{eq-gslintsep}
\end{equation}
For example, if a matter system with entropy ${\cal S}$ enters a black hole, then 
$S_\text{out}[\sigma_2]-S_\text{out}[\sigma_1]={\cal S}$, so by Eq.~(\ref{eq-gslintsep}) the horizon area must increase at least by $4G\hbar {\cal S}$.

\subsection{Quantum Focussing Conjecture and Bousso Bound}
\label{sec-qfc}

Returning to a general null hypersurface $H$, we can consider how the quantum expansion, in turn, varies under second deformations of the cut $\sigma$. The {\em Quantum Focussing Conjecture} (QFC) states that the quantum expansion at $y$ will not increase under a deformation of the cut at $\bar y$~\cite{BouFis15a}. The second deformation at $\bar y$ is required to be taken in the same direction (future or past) with respect to which the quantum expansion was defined at $y$.

For $y\neq \bar y$, the QFC can be proven using strong subadditivity. Below we will focus on the most nontrivial case, $y=\bar y$. Then the QFC can be stated as
\begin{equation}
\Theta'[\sigma;y]\leq 0~,
\label{eq-qfc}
\end{equation}
using the notation introduced around Eq.~(\ref{eq-qe}). Substituting Eq.~(\ref{eq-sgendef}) yields the separated differential form of the QFC
\begin{equation}
-\frac{\theta'}{4G\hbar} \geq \frac{1}{\mathcal A} \left(S''_\text{out}-S_\text{out}'\theta\right)~,
\label{eq-qfcsep}
\end{equation}
where we have suppressed the dependence of all quantities on $[\sigma;y]$.

Now consider two cuts $\sigma_1$ and $\sigma_2$ of $H$. Suppose that the cut $\sigma_2$ has larger or equal $w$ on every generator, and that $\Theta[\sigma_1;y]\leq 0$ at every $y$ where the cuts differ. Then integration of Eq.~(\ref{eq-qfc}) implies that the quantum expansion remains nonpositive between $\sigma_1$ and $\sigma_2$. A second integration yields
\begin{equation}
S_\text{gen}[\sigma_2]\leq S_\text{gen}[\sigma_1]~.
\label{eq-qfcint}
\end{equation}

This looks the same as the GSL; and indeed the above argument can be applied to the special case of causal horizons. Under the (physically reasonable) assumption that their quantum expansion vanishes at late times, integrating the QFC once (towards the past) implies the differential version of the GSL, Eq.~(\ref{eq-gsl}). Integrating the QFC twice implies the integral version of the GSL, Eq.~(\ref{eq-gslint}).

However, the GSL does not imply the QFC on causal horizons, so the QFC is stronger. Further, the QFC is more general, since Eqs.~(\ref{eq-qfc}) and (\ref{eq-qfcint}) apply to arbitrary null hypersurfaces. In this general setting, Eq.~(\ref{eq-qfcint}) can be regarded as a quantum-corrected version of the Bousso bound~\cite{BouFis15a}. (It is distinct from the QBB discussed in Sec.~\ref{sec-casini}~\cite{BouCas14a}.) The assumption that $\Theta[\sigma_1;y]\leq 0$ together with Eq.~(\ref{eq-qfc}) is the quantum generalization of the defining condition for light-sheets, that $\theta\leq 0$ everywhere between $\sigma_1$ and $\sigma_2$. Upon separating the area and matter entropy terms in the integrated QFC, one obtains
\begin{equation}
S_\text{out}[\sigma_2]-S_\text{out}[\sigma_1] \leq \frac{A[\sigma_1]-A[\sigma_2]}{4G\hbar} ~.
\label{eq-qfcintsep}
\end{equation}
For well-isolated matter systems localized to the light-sheet between $\sigma_1$ and $\sigma_2$ the left hand side can be identified as the entropy ${\cal S}$ on the light-sheet~\cite{BouFis15a}, and one recovers the Bousso bound~\cite{CEB1,FMW},
\begin{equation}
{\cal S}\leq \frac{A[\sigma_1]-A[\sigma_2]}{4G\hbar} ~.
\end{equation}

\section{Standard Weak Gravity Limit}
\label{sec-weak}

We now review the weak gravity limit of the GSL~\cite{Cas08,Wal10,Wal11} and the QFC~\cite{BouFis15a,BouFis15b}. One obtains two nongravitational statements, i.e., statements about quantum field theory. Both have been proven for free fields~\cite{Wal11,BouFis15b}.  In addition we will review a third statement that has been formulated only in this limit, a bound on the vacuum-subtracted entropy of a bounded region~\cite{BouCas14a,BouCas14b}.  This statement has been proven for free and interacting theories.

\subsection{Focussing in the $G\to 0$ Limit}
\label{sec-weakfocus}

Any null hypersurface $H$ is ruled by a congruence of null geodesics, its {\em generators}. Given the expansion at one point on $H$, the expansion at any other point on the same generator can be computed by integrating the Raychaudhuri equation:
\begin{equation}
  \theta'  = -\frac{\theta^2}{2} -\varsigma_{ab}\varsigma^{ab} - 8\pi G\, T_{ww}~.
\label{eq-raych}
\end{equation}
Here 
\begin{equation}
T_{ww}\equiv \langle T_{ab} k^ak^b\rangle~,
\label{eq-twwkk}
\end{equation}
$T_{ab}$ is the stress tensor, $k^a = dx^a/dw$ is the null vector tangent to the generator, and $\theta = \nabla_ak^a$; see also Eq.~(\ref{eq-thetadef}). The shear $\varsigma_{ab}$ is the traceless part of the null extrinsic curvature:
\begin{equation}
\varsigma_{ab}[\sigma] \equiv q_a^{~c} q_b^{~d}\, \nabla_c k_d 
-\frac{1}{2} \theta \tilde q_{ab} ~,
\end{equation}
where $q_{ab}$ is the intrinsic metric on the cut $\sigma$. 

In the weak gravity limit, one considers a two-dimensional spatial surface $\sigma_1$ whose null expansion vanishes at all points $y$ at leading order:
\begin{equation}
\theta[\sigma_1;y]  = 0 + O(G)~.
\label{eq-theta0}
\end{equation}
The goal is to compute the expansion elsewhere on the null hypersurface $H$ of which $\sigma_1$ is a cut. We will also assume that $\varsigma_{ab}$ is at most of order $G^{1/2}$ on $H$.
Thus, $\theta$ will be generated only at order $G$; the $\theta^2$ term in Eq.~(\ref{eq-raych}) will be $O(G^2)$ and can be neglected:
\begin{equation}
  \theta'  = -8\pi G\, {\cal T} + O(G^2) ~,
\label{eq-weakraych}
\end{equation}
where we have defined
\begin{equation}
{\cal T}\equiv T_{ww}+\tilde\varsigma^2~,~~ \tilde\varsigma^2\equiv \frac{\varsigma_{ab}\varsigma^{ab}}{8\pi G}~.
\end{equation}
One can then compute the expansion $\theta$ on any other cut $\sigma_2$ of $H$ by direct integration of Eq.~(\ref{eq-raych}) along the generator $y$ on which the point lies:
\begin{eqnarray} 
\theta[\sigma_2;y] &  = & \theta[\sigma_1;y] - \nonumber
8\pi G \int_{\sigma_1}^{\sigma_2}\!\!\! dw~ {\cal T} \\
& + & 
O(G^2) ~,
\label{eq-weaktheta}
\end{eqnarray} 
where the integral runs over the single generator $y$.  This is the key result for focussing in the weak gravity limit.

Integrating a second time yields a formula for the area change accumulated along the generator $y$. This can be integrated over all generators to yield the area difference between the two cuts. For simplicity we quote the result for the case where $\theta[\sigma_1;y]$ vanishes through $O(G)$: 
\begin{eqnarray} 
A[\sigma_1] & - & A[\sigma_2] = \int_{\sigma_1}^{\sigma_2}\! d^2y\sqrt{h(y)}\, dw \, \theta(w,y) \nonumber \\
& = & 8\pi G \int_{\sigma_2}^{\sigma_1}\! d^2y\sqrt{h(y)}\, dw \, \left[w-w_2(y)\right]\, {\cal T}\nonumber\\
& & + O(G^2)~,
\label{eq-weakarea}
\end{eqnarray} 
where the integral now runs over the entire portion of $H$ between the two cuts. The second line follows from Eq.~(\ref{eq-weaktheta}) and integration by parts; $w_2(y)$ is the value of $w$ on $\sigma_2$ at $y$.

\subsection{Weak Gravity Generalized Second Law}
\label{sec-weakgsl}

We now apply the above results to the cut $\sigma$ of a causal horizon $H$. With $\sigma_2=\sigma$ and $\sigma_1\to \infty$, substituting  Eq.~(\ref{eq-weaktheta}) into Eq.~(\ref{eq-gslsep}) yields the weak gravity limit of the GSL:
\begin{equation}
-\frac{S_\text{out}'[\sigma;y]}{\cal A} \leq \frac{2\pi}{\hbar} \int_{\sigma}^\infty  dw ~{\cal T}~,
\label{eq-weakgsl}
\end{equation}
where the integral runs over the single generator $y$.

We have taken the limit as $G\to 0$. The GSL remains nontrivial in this limit, because the leading factor of $G$ in Eq.~(\ref{eq-weaktheta}) cancels against the $G$ in the denominator in Eq.~(\ref{eq-gsl}). Thus the GSL reduces to an exact statement concerning the von Neumann entropy of quantum fields restricted to a semi-infinite portion of a causal horizon.

The use of both Eq.~(\ref{eq-weaktheta}) and Eq.~(\ref{eq-gslsep}) requires justification. Eq.~(\ref{eq-weaktheta}) is valid only if the cut $\sigma_1$ has vanishing expansion at order $G^0$. Here, that surface is taken to be in the infinite future on a causal horizon~\cite{Haw71}, where $\theta$ indeed vanishes. Eq.~(\ref{eq-gslsep}) is the ``separated'' differential version of the GSL. This version requires us to separately control the divergences in the entanglement entropy, and the RG flow of Newton's constant. But in the weak gravity limit, we have seen that Newton's constant cancels out. The divergent boundary contribution $S_\text{out}$ also drops out, because only $S_\text{out}'$ enters, and at leading order, the derivative is taken along a null hypersurface with fixed cross-sectional geometry. Hence $S_\text{out}'$ in Eq.~(\ref{eq-weakgsl}) is well-defined.

Integration by parts of Eq.~(\ref{eq-weakgsl}), or substitution of Eq.~(\ref{eq-weakarea}) into Eq.~(\ref{eq-gslintsep}), yields the integrated version of the weak gravity GSL:
\begin{equation}
S_\text{out}[\sigma_2]-S_\text{out}[\infty]\leq
\frac{2\pi}{\hbar}\!\int_{\sigma_2}^{\infty}\!\!\! d^2y \sqrt{h(y)}\, dw\left[w-w_2(y)\right]{\cal T}.
\label{eq-weakgslint}
\end{equation}
The upper limit $\infty$ can in practice be replaced by any cut $\sigma_1$ (to the future of $\sigma_2$) on which the expansion vanishes or makes a negligible contributions to the area difference.

\subsection{Quantum Null Energy Condition}
\label{sec-qnec}

The QFC, too, becomes separable into area and exterior entropy terms as $G\to 0$. Thus we may use Eq.~(\ref{eq-qfcsep}) as we study the weak gravity limit of the QFC. Substituting Eq.~(\ref{eq-weakraych}) and taking $G\to 0$ yields the {\em Quantum Null Energy Condition} (QNEC):
\begin{equation}
\frac{S_\text{out} ''[\sigma;y]}{\cal A} \leq \frac{2\pi}{\hbar}\, {\cal T}~.
\label{eq-qnec}
\end{equation}
The QNEC holds on any generator $y$ orthogonal to a slice $\sigma$ with the property that $\theta[\sigma;y]$ scales as a  positive power of $G$ as the limit is taken, or else the $S_\text{out}' \theta$ term in Eq.~(\ref{eq-qfcsep}) will contribute. This will be the case everywhere on $H$ if $H$ is a causal horizon. Thus, integration of the QNEC, Eq.~(\ref{eq-qnec}), implies the weak-gravity GSL, Eq.~(\ref{eq-weakgsl}).

More generally, given any point $p$ and null vector $k$ at $p$ in an arbitrary spacetime, one can find a spatial surface with null normal vector $k$ and vanishing expansion in an open neighborhood of $p$. Taking $H$ to be the null hypersurface orthogonal to any Cauchy-splitting completion $\sigma$ of this surface, Eq.~(\ref{eq-qnec}) applies. Note that $S_\text{out} ''$ will in general depend on the choice of $\sigma$.

\subsection{Quantum Bousso Bound}
\label{sec-casini}

The {\em Quantum Bousso Bound} (QBB) was formulated and proven in Refs.~\cite{BouCas14a,BouCas14b}. Let $H$ be a (classical) light-sheet~\cite{CEB1} in the weak gravity limit. That is, we assume that $H$ is a null hypersurface bounded by cuts $\sigma_1$, $\sigma_2$ such that $\theta$ is nonpositive and at most $O(G)$, everywhere between $\sigma_1$ and $\sigma_2$. Then
\begin{equation}
S_C\leq \frac{A[\sigma_1]-A[\sigma_2]}{4G\hbar}~,
\label{eq-qbb}
\end{equation}
where $S_C$ is the vacuum-subtracted entropy or Casini entropy~\cite{MarMin04,Cas08} of the quantum state restricted to $H$.

The bound is tied to the weak gravity limit, because $S_C$ is well-defined only as $G\to 0$. As backreaction gets small, different quantum states become compatible with the same spacetime geometry, that of $H$. Then it is possible to restrict both an arbitrary state $\rho_{\rm global}$, and the vacuum state $|0\rangle \langle 0|$, to $H$. This yields reduced density operators $\rho$ and $\rho_0$, with von Neumann entropies $S[\rho]$ and $S[\rho_0]$. One defines
\begin{equation}
S_C\equiv S[\rho]-S[\rho_0]~.
\end{equation}

Like the QFC, the QBB also reduces to the original covariant bound~\cite{CEB1,FMW} in settings where systems are well-isolated. However, the QBB has no known extension to strongly gravitating regions. The QBB is not known to imply, nor to follow from, any other entropy bounds listed above.

In the earlier subsections, we expressed the weak-gravity limit of entropy bounds by converting (derivatives of) the area to expressions involving the energy flux, by applying Eqs.~(\ref{eq-weakraych}) and (\ref{eq-weaktheta}). We could also convert Eq.~(\ref{eq-qbb}) in this way; by using Eq.~(\ref{eq-weakarea}). Terms proportional to $\theta_0$ would have to be restored in Eq.~(\ref{eq-weakarea}), as they may be necessary to uphold the classical nonexpansion condition~\cite{BouCas14a}. Thus, we would obtain an expression similar in form to the integrated GSL, but for a different entropy and with the integral running over finite affine distance.

However, it is possible to write the QBB in a stronger form,
\begin{equation}
S_C\leq \frac{2\pi}{\hbar}\Delta K~,
\label{eq-sdk}
\end{equation}
where $\Delta K$ is the vacuum-subtracted modular energy of $\rho$ on $H$~\cite{BouCas14a,BouCas14b}. One can show that this implies Eq.~(\ref{eq-qbb}). Yet $\Delta K$, like $\Delta A$, can be expressed as an integral over the energy flux in the weak gravity limit. Since this is the format we seek, it makes sense to start from the stronger statement, Eq.~(\ref{eq-sdk}). 

In general the modular energy is highly nonlocal, but for the finite light-sheets $H$, one has\footnote{The $\tilde\varsigma^2$ term enters the modular Hamiltonian through the effective stress tensor of gravitational radiation~\cite{Wal11}.} 
\begin{equation}
\Delta K =  \int_{\sigma_1}^{\sigma_2} d^2y \sqrt{h(y)} \,dw\, g_y(w)\,{\cal T}~,
\label{eq-dk}
\end{equation}
where
\begin{equation}
g_y(w) = \frac{(w-w_1)(w_2-w)}{w_2-w_1}
\label{eq-gfree}
\end{equation}
for free theories. The weight function $g$ depends on $y$ through $w_1$ and $w_2$, the affine parameter values where the generator $y$ intersects $\sigma_1$ and $\sigma_2$. One can show that Eqs.~(\ref{eq-sdk}) and (\ref{eq-dk}) imply Eq.~(\ref{eq-qbb}). Moreover, Eq.~(\ref{eq-sdk}) follows directly from the positivity of the relative entropy. 

For interacting theories $g(w)$ will be a different function that satisfies constraints derived in Ref.~\cite{BouCas14b}. Under changes of the affine parameter, $g$ transforms nontrivially due to renormalization:
\begin{equation}
g(w)\to g(u) = g(w(u)) \frac{du}{dw}~.
\label{eq-gg}
\end{equation}
The constraints on $g$ are sufficient to obtain Eq.~(\ref{eq-qbb}) from Eq.~(\ref{eq-sdk}) and (\ref{eq-dk}).  

\section{Asymptotic Weak Gravity Limit}
\label{sec-asymptotic}

In an asymptotically flat spacetime, gravity effectively becomes weak for light-rays near future null infinity, ${\cal I}^+$, for any fixed value of $G$. This is because the relevant term in the focussing equation is $8\pi G {\cal T}$. One way to make this term small is to take $G\to 0$ at fixed ${\cal T}$. But another is to take the energy density ${\cal T}\to 0$ at fixed $G$. This is precisely what happens near ${\cal I}^+$. Only massless fields reach ${\cal I}^+$, and their density scales as ${\cal T}\sim O(r^{-2})$ at large distances. In this section, we will reconsider the weak gravity limit as an expansion in $G/r^2$, with $G$ fixed and $r\to \infty$.

We will be able to work strictly on a Minkowski background; we will never need to consider any other metric explicitly. The effects of gravity can be computed at order $G/r^2$ by integrating the evolution equations for the null extrinsic curvature (e.g., the Raychaudhuri equation for the expansion), along a null congruence.

\subsection{Causal Horizons in Asymptotically Flat Space}
\label{sec-hup}

The metric of Minkowski space can be written as
\begin{eqnarray} 
ds^2 
& = &  -du^2 -2 du\, dr + r^2 d\Omega^2 \label{eq-ur}\\
& = &  -dw^2 -2 dw\, dz + d\rho^2 +\rho^2 d\phi^2 \label{eq-wz}~.
\end{eqnarray}
In terms of the standard spherical and Cartesian systems, $u=t-r$, $w=t-z$, 
and $\rho = (x^2+y^2)^{1/2} = r \sin\vartheta$. We use the notation 
\begin{equation}
\Omega = (\vartheta,\phi)~,~~d\Omega^2 = d\vartheta^2 + \sin^2\vartheta\, d\phi^2~.
\end{equation}

In the limit as $r\to \infty$, the coordinates $(u,\Omega)$ label points on future null infinity, ${\cal I}^+$. In the standard ``unphysical spacetime'' or Penrose diagram of Minkowski space~\cite{HawEll,Wald}, ${\cal I}^+$ is a null hypersurface ruled by null geodesics with affine parameter $u$. Each null geodesic is labelled by its angular position $\Omega$. 

We now consider a general asymptotically flat spacetime $M$~\cite{Wald}. It has the same conformal boundary as Minkowski space, and we shall continue to use coordinates $(u,\Omega)$ on ${\cal I}^+$. Let $p\in {\cal I}^+$ be a point at affine time $u_p$ and angle $\Omega_p$. Without loss of generality we set $\vartheta_p=\pi$. Let $H(u_p)$ be the boundary of the past of $p$:
\begin{equation}
H(u_p)\equiv \dot I^-(p)~,~~p\in {\cal I}^+ ~.
\label{eq-hdef}
\end{equation}
More precisely, $H$ is the hypersurface in $M$ obtained by finding $\dot I^-(p)$ in the unphysical spacetime and then transforming back to $M$.

In pure Minkowski space, $H(u_p)$ is the null plane~\cite{HawEll} given by $t+z= u_p$, or
\begin{equation}
w+2z=u_p~.
\label{eq-hupmink}
\end{equation}
This is a null hypersurface with no expansion and shear. It is ruled by light-rays with affine parameter $w$. Each geodesic is labelled by its position on the transverse $(x,y)$ or ($\rho,\phi$) plane.

Now consider $H(u_p)$ in a general asymptotically flat spacetime $M$. Because $H(u_p)$ is the boundary of the past of a set, it is still a null hypersurface~\cite{HawEll,Wald}. This implies that the QFC (Sec.~\ref{sec-qfc}) can be applied to $H(u_p)$, and hence, Eqs.~(\ref{eq-qfc}) and (\ref{eq-qfcint}) apply to $H(u_p)$. More strongly, $H(u_p)$ is a causal horizon, because $p$ is the endpoint of an inextendible worldline (e.g., an accelerated timelike observer, or a light-ray ending at $p$). This implies that the GSL, Eqs.~(\ref{eq-gsl}) and (\ref{eq-gslint}), can be applied to $H(u_p)$.

So far, no limits have been taken or approximations made. Near $p$ (i.e., for sufficiently large $w$), $H(u_p)$ will resemble the null plane of Minkowski space, Eq.~(\ref{eq-hupmink}).  For general $u_p$, $H(u_p)$ may suffer significant distortions deep inside $M$, as it passes through strongly gravitating regions. 

\subsection{Late Causal Horizons as a Weak Gravity Limit}
\label{sec-limit}

We now take the limit as $u_p\to \infty$. That is, we take the endpoint of the causal horizon $H(u_p)$ to approach future timelike infinity. For simplicity, we will assume that all energy is eventually radiated out to ${\cal I}^+$; that is, the Bondi mass vanishes at sufficiently late retarded times. Thus, for sufficiently large $u_p$, the causal horizon $H(u_p)$ will remain outside all timelike matter sources, passing through the radiative region only (see Fig.~\ref{fig-zones}). This region is filled with radiation that propagates to ${\cal I}^+$ along radially outgoing light-rays.

To be precise, let us pick an early time cutoff $u_E(\Omega)$ of ${\cal I}^+$ and exclude the portion of each $H(u_p)$ that lies in the past of $u_E$ from consideration. Then we take the limit as $u_p\to \infty$. Physically this is not restrictive, since one can choose $u_E$ to be as early as we like, e.g., prior to the arrival of the earliest radiation at ${\cal I}^+$. This prescription excludes regions where $H(u_p)$ would develop caustics. Similarly, we may place cutoffs on ${\cal I}^-$ such that no radiation enters the spacetime before the early time cutoff and after the late-time cutoff (up to weak tails~\cite{Chr91,ChrKla93}). This ensures that the above assumptions about ${\cal I}^+$ can be satisfied.

For large enough $u_p$, the energy flux through $H(u_p)$ is dominated by radially outgoing radiation. This implies that $T_{ww}\sim 1/u_p^2$, as we will explain in detail below. The Weyl tensor sources the shear $\varsigma$ of $H(u_p)$. Since the relevant component of the Weyl tensor falls off as $1/u_p$, we have $\tilde \varsigma^2\sim u_p^{-2}$~\cite{BouKoe16}. Hence 
\begin{equation}
{\cal T}\sim u_p^{-2}~,
\end{equation}
so the focussing equation,
\begin{equation}
  \theta'  = -\frac{\theta^2}{2} -8\pi G\, {\cal T}~,
\end{equation}
implies that the expansion $\theta$ is sourced at order $1/u_p^2$. Therefore, the $\theta^2$ term can be neglected in the $u_p\to\infty$ limit, just as it could in the $G\to 0$ limit in Sec.~\ref{sec-weak}. 

Thus, the weak gravity focussing equation (\ref{eq-weakraych}) holds on $H(u_p)$ for large $u_p$, even though we have not taken $G\to 0$. Also, the weak gravity equations for the expansion and for the area difference, Eqs.~(\ref{eq-weaktheta}) and (\ref{eq-weakarea}), both hold on $H(u_p)$ for large $u_p$.

It follows that the weak gravity entropy bounds reviewed in Sec.~\ref{sec-weak} can all be applied directly to $H(u_p)$: the differential and integral form of the weak gravity GSL, Eqs.~(\ref{eq-weakgsl}) and (\ref{eq-weakgslint}); the QNEC, Eq.~(\ref{eq-qnec}); and the QBB, Eq.~(\ref{eq-qbb}). 

We stress that we have not formulated any new entropy bounds. Rather, we have identified $H(u_p)$, for large enough $u_p$, as a null surface to which these known bounds apply. Our remaining task is to take the $u_p\to\infty$ limit and express the bounds directly in terms of variables on ${\cal I}^+$.

\section{Entropy Bounds on ${\cal I}^+$}
\label{sec-scri}

We have constructed null hypersurfaces near ${\cal I}^+$ to which known weak-gravity entropy bounds apply directly. To make this explicit, we referred only to bulk quantities such as the physical energy flux and shear tensor. 

In this section, we express our results in terms of a finite boundary energy flux (in which the role of the shear is played by the Bondi news). In particular, our results will show that the bounds obtained from the null surfaces $H(u_p)$ are independent of their orientation in the spacetime, i.e., independent of which boundary generator the point $p$ lies on.

Like in the previous section, it suffices to carry out all calculations strictly on a Minkowski background. This simplifies the analysis considerably.

\subsection{Bulk to Boundary Dictionary}

Before taking the $u_p\to\infty$ limit, we have to establish relations between bulk and boundary quantities. Below, $u_p$ and $r$ are taken to be large, but still finite. Boundary quantities are denoted by hats.

\paragraph{Cuts}
All of the entropy bounds involve specifying a cut on a null surface, and then either a deformation of this cut or a second cut. In order to derive entropy bounds on ${\cal I}^+$, we need to map cuts and deformations specified on ${\cal I}^+$ into cuts and deformations on the bulk null surfaces $H(u_p)$, for large $u_p$.

Let $\hat\sigma$ be a cut on ${\cal I}^+$.  For large enough $u_p$, the boundary of the past of $\hat\sigma$ defines a cut $\sigma^{(u_p)}$ of each $H(u_p)$:
\begin{equation}
\sigma^{(u_p)} \equiv H(u_p)\cap \dot I^-(\hat\sigma)~.
\label{eq-sigmadef}
\end{equation}
This definition is appropriate because the cuts ``flow with the outgoing radiation''. That is, the physical radiation that lies above (or below) the cut $\sigma^{(u_p)}$ on $H(u_p)$ becomes independent of $u_p$ for large $u_p$. Hence, the entropy of the quantum state between two cuts, or on the semi-infinite region on one side of a cut, will become independent of $u_p$ in the limit.

A manifestly local definition equivalent to Eq.~(\ref{eq-sigmadef}) is to associate to each point on the boundary the bulk points with the same $(u,\Omega)$ but varying $r\leq \infty$. This implies that we associate to each $(u,\vartheta,\phi)$ on ${\cal I}^+$ the point $(w,\rho,\phi)$ on $H(u_p)$, with
\begin{eqnarray} 
  w & = & (u_p-u) \tan^2\frac{\vartheta}{2} + u~, \label{eq-wutheta}\\
  \rho & = & (u_p-u)\tan\frac{\vartheta}{2}~. \label{eq-rhoutheta}
\end{eqnarray} 
In particular, if $\hat\sigma$ is given as a function $u(\Omega)$, the above equations determine the cut $\sigma^{(u_p)}$ parametrically as $w(\Omega)$, $\rho(\Omega)$.  

For completeness we note that $z=(u_p-w)/2$ and
\begin{equation}
r = \frac{u_p-u}{2\cos^2 (\vartheta/2)}~.
\label{eq-rutheta}
\end{equation}
The $r$-position of the cut is irrelevant to the entropy $S_\text{out}[\sigma^{(u_p)}]$, since the radiation is propagating radially outward as $u_p\to \infty$. Note that $u_p-u$ can be assumed positive since the limit $u_p\to\infty$ is taken with $(u,\theta,\phi)$ fixed.

\paragraph{Area Element}
We must also characterize local deformations of $\sigma^{(u_p)}$. As $u_p\to\infty$, they should limit to a deformation of the cut $\hat\sigma$, whereby a small area element of size $\delta\Omega$ is pushed forward along the null generator of ${\cal I}^+$ at angular position $\Omega$. This can be accomplished by locating the generator of $H(u_p)$ that has the same angular position on $\sigma^{(u_p)}$, using Eq.~(\ref{eq-rhoutheta}). The solid angle element $\delta\Omega$ spans an area element of size
\begin{equation}
{\cal A} = r^2 \delta\Omega
\label{eq-calaomega}
\end{equation}
on $\sigma^{(u_p)}$. Note that this depends on $u_p$ and $\vartheta$ only through Eq.~(\ref{eq-rutheta}).

Eq.~(\ref{eq-calaomega}) follows immediately from setting $du=0$ in Eq.~(\ref{eq-ur}), but the result may seem counterintuitive. ${\cal A}$ is the area of the intersection of a small solid angle $\delta \Omega$ of a light-cone $u=\mbox{const}$ with a null plane. In Euclidean space, the intersection area of an angle element of a cone with a plane would depend not only on the plane's distance $r$ from the apex, but also on the angle at which the cone and plane meet. Here the latter dependence is absent for small $\delta\Omega$, because the cone is null. 

\paragraph{Null Tangent Vector}
We consider the null surface $H(u_p)$ with affine parameter $w$. The null vector $k^\mu=dx^\mu/dw$ is tangent to its generators. $k^\mu$ has components $(0,1,0,0)$ in the $\{u_p, w, \rho, \phi\}$ coordinate system of Minkowski space. In the $\{u,r,\vartheta,\phi\}$ coordinates, $k^\mu$ has components 
\begin{eqnarray} 
k^u = (du/dw)_{u_p,\rho,\phi} & = & \cos^2(\vartheta/2)~,
\label{eq-ku} \\
k^r = (dr/dw)_{u_p,\rho,\phi} & = & -(\cos\vartheta)/2~,
\label{eq-kr}\\
k^\vartheta = (d\vartheta/dw)_{u_p,\rho,\phi} & = & \frac{\sin\vartheta\cos^2(\vartheta/2)}{u_p-u}~,
\label{eq-ktheta}\\
k^\phi = (d\phi/dw) _{u_p,\rho,\phi} 
\label{eq-kphi} & = & 0~.
\end{eqnarray}
The subscripts indicate coordinates that are held fixed as the derivative is taken. 

\paragraph{Affine Parameter and Angular Dependence}
We are interested in following the same infinitesimal beam of radiation near fixed $u,\vartheta,\phi$ through a sequence of light-sheets $u_p=$const as we take $u_p\to\infty$. This is why we expressed $k^\mu$ in terms of $u,\vartheta,\phi$. The details of the $u$ and $\vartheta$ dependence will not be important. What matters is the scaling with $u_p$, or equivalently (to leading order at fixed $u,\vartheta,\phi$) by Eq.~(\ref{eq-rutheta}), the scaling with $r$.

Because $d\vartheta/dw$ falls off as $u_p^{-1}$, $u$ becomes an affine parameter on $H(u_p)$ as $u_p\to\infty$:
\begin{equation}
\frac{d^2 u}{dw^2}=-\frac{\sin\vartheta}{2}\left(\frac{d\vartheta}{dw}\right)_{u_p,\rho,\phi} \sim O(u_p^{-1})~.
\label{eq-secondder}
\end{equation}
This implies that on a fixed generator $(\rho,\phi)$ of $H(u_p)$ and for finite $w_1,w_2$,
\begin{equation}
\lim_{u_p\to\infty} u(w_2)-u(w_1) = (w_2-w_1)\left(\frac{du}{dw}\right)_{u_p,\rho,\phi}~,
\label{eq-wuwu}
\end{equation}
where the last factor can be evaluated anywhere between $w_1$ and $w_2$. The scaling of $d\vartheta/dw$ also implies that integrals over a generator of $H(u_p)$ become integrals over a boundary generator at fixed angular position $\Omega$, in the $u_p\to\infty$ limit. (This assumes that either the integral has finite range of $u$, or the integrand drops off sufficiently rapidly at large $u$.)

\paragraph{Stress Tensor and Shear}

A finite boundary ``matter stress tensor'' on ${\cal I}^+$ (really, the nongravitational energy flux across ${\cal I}^+$) can be defined as~\cite{StrZhi14,FlaNic15}:
\begin{eqnarray} 
\hat T_{uu}(u,\vartheta,\phi) \equiv \lim_{r\to \infty} r^2\, T_{uu}(u,r,\vartheta,\phi)~.
\label{eq-hattuudef}
\end{eqnarray} 
We now relate this quantity to $T_{ww}$, the nongravitational energy flux across $H(u_p)$.

The bulk matter stress tensor in Minkowski space~\cite{FlaNic15}, in the $\{u,r,\vartheta,\phi\}$ coordinates, falls off as
\begin{eqnarray}
T_{uu} & \propto & r^{-2}~,\\
T_{u\vartheta} & \propto & r^{-2}~,\\
T_{\vartheta\vartheta} & \propto & r^{-1}~.
\end{eqnarray} 
Other relevant components fall off at least as rapidly as $r^{-3}$. These fall-off conditions together with Eqs.~(\ref{eq-twwkk}) and (\ref{eq-ku}-\ref{eq-kphi}) yield
\begin{eqnarray} 
T_{ww} & = & \frac{\hat T_{uu}}{r^2} \left(\frac{du}{dw}\right)_{u_p,\rho,\phi}^2 + O(r^{-3}) \\
& = & \frac{\hat T_{uu}}{r^2} \cos^4\frac{\vartheta}{2}+ O(r^{-3})~.
\label{eq-twwdrop}
\end{eqnarray} 

This result implies that the boundary nongravitational flux $\hat T_{uu}$ could have been defined directly as the $u_p\to\infty$ limit of $r^2 T_{ww}/\cos^2(\vartheta/2)$. Given the appearance of $\tilde\varsigma^2$ alongside $T_{ww}$ in the focussing equation for $H(u_p)$, it is natural to define a  {\em boundary shear} as a limit of the rescaled shear of the null surfaces $H(u_p)$:
\begin{equation}
\hat\varsigma_{ab}(u,\vartheta,\phi) \equiv \frac{1}{\sqrt{8\pi G}}\lim_{r\to\infty} r\, \frac{\varsigma_{ab} (u,r,\vartheta,\phi)}{\cos^2(\vartheta/2)}~.
\label{eq-hatsigmadef}
\end{equation}
Indeed, it can be shown~\cite{BouKoe16} that $\hat\varsigma_{ab}=-N_{ab}/2$, where $N_{ab}$ is the Bondi news. This confirms that $\hat\varsigma_{ab}$, like $\hat T_{uu}$, is finite and independent of the orientation of $H(u_p)$. 
We conclude that
\begin{equation}
\tilde\varsigma^2 = \frac{\hat\varsigma_{ab}\hat\varsigma^{ab}}{r^2}  \cos^4\frac{\vartheta}{2} + O(r^{-3})~.
\label{eq-ssdrop}
\end{equation}

The total boundary energy flux is defined as
\begin{equation}
\hat{\cal T}\equiv \hat T_{uu} + \hat\varsigma_{ab}\hat\varsigma^{ab}~.
\end{equation}
By Eqs.~(\ref{eq-twwdrop}) and (\ref{eq-ssdrop}), the total energy flux across $H(u_p)$ is given by
\begin{equation}
{\cal T}(u_p,u,\Omega) = \frac{\hat{\cal T}(u,\Omega)}{r^2} \cos^4\frac{\vartheta}{2} + O(r^{-3})~,
\label{eq-t}
\end{equation} 
where $r(u_p,u,\Omega)$ is given by Eq.~(\ref{eq-rutheta}). This is a central result for what follows.

\subsection{Quantum Null Energy Condition on ${\cal I}^+$}

We showed in Sec.~\ref{sec-weak} that the standard QNEC applies to the null surfaces $H(u_p)$ for large $u_p$. Eq.~(\ref{eq-qnec}) can be written as
\begin{equation}
\frac{1}{\cal A} S_\text{out}''[\sigma^{(u_p)};y(\Omega)] \leq \frac{2\pi}{\hbar}~ {\cal T}~.
\end{equation}
Here $S_\text{out}[\sigma^{(u_p)}]$ is the entropy of the quantum state on $H(u_p)$ restricted to one side of the cut $\sigma^{(u_p)}$. $y(\Omega)$ marks the generator of $H(u_p)$ that has angular position $\Omega$ on $\sigma^{(u_p)}$; this is where the cut is varied. $\sigma^{(u_p)}$ is defined in terms of a cut $\hat\sigma$ on ${\cal I}^+$ through Eqs.~(\ref{eq-wutheta}) and (\ref{eq-rhoutheta}).

Using Eqs.~(\ref{eq-calaomega}) and~(\ref{eq-t}), this becomes
\begin{eqnarray} 
\frac{1}{r^2\delta \Omega} \frac{du}{dw} \frac{d}{du}\left(\frac{du}{dw}\frac{d}{du} 
S_\text{out}[\sigma^{(u_p)},\Omega] \right) & & \nonumber \\ 
\leq \frac{2\pi}{\hbar} \frac{\hat{\cal T}(\Omega)}{r^2} \left(\frac{du}{dw}\right)^2 + O(r^{-3})~. & & 
\end{eqnarray} 
The product rule yields two terms, of which one is proportional to $d^2u/dw^2$. This is subleading by Eq.~(\ref{eq-secondder}) and can be dropped along with other terms of order $r^{-3}$ as we take the limit $u_p\to\infty$. Upon cancelling the factor $r^{-2}(du/dw)^2$ on both sides we obtain
\begin{equation}
\frac{1}{\delta\Omega} \frac{d^2}{du^2} \hat S_\text{out}[\hat\sigma,\Omega] \leq \frac{2\pi}{\hbar} \hat{\cal T}(\Omega)~,
\label{eq-qnecscri}
\end{equation}
where $\hat S_\text{out}[\hat\sigma]$ is the entropy of the reduced density operator obtained by restricting the quantum state on ${\cal I}^+$ to the past\footnote{At any finite value of $G,u_p$, $H(u_p)$ encounters caustics in the distant past, far from ${\cal I}^+$. Strictly speaking, this implies that our first order treatment in $G/r^2$ breaks down when $S_\text{out}$ refers to the past side. This can be resolved by replacing the surfaces $H(u_p)$ with $\bar H(v_q)$ in the above derivation. Here $\bar H(v_q)$ is the boundary of the future of a point $q$ on ${\cal I}^-$ that approaches spatial infinity, $v_q\to\infty$.} or future of the cut $\hat\sigma$. This is the QNEC on ${\cal I}^+$. 

\subsection{Generalized Second Law on ${\cal I}^+$}

Integration of the QNEC on ${\cal I}^+$ over the single generator $\Omega$ yields the differential form of the GSL on ${\cal I}^+$:
\begin{equation}
-\frac{1}{\delta\Omega} \frac{d}{du}\hat S_\text{out}[\hat\sigma;\Omega] \leq 
\frac{2\pi}{\hbar}\int_{\hat\sigma}^\infty du~ \hat{\cal T}~.
\label{eq-gslscri}
\end{equation}

We can also derive Eq.~(\ref{eq-gslscri}) directly from the GSL in the bulk. The argument mirrors our derivation of the boundary QNEC. 
The only new feature is the appearance of the integral over a generator of $H(u_p)$ in the bound. By the remarks following Eq.~(\ref{eq-wuwu}), this becomes an integral over the boundary generator $\Omega$ in the limit.

Integration by parts of Eq.~(\ref{eq-gslscri}) yields the integral form of the GSL on ${\cal I}^+$:
\begin{equation}
\hat S_\text{out}[\hat\sigma_2]-\hat S_\text{out}[\infty]\leq
\frac{2\pi}{\hbar}\!\int_{\hat\sigma_2}^{\infty}\!\!\! d^2\Omega du 
[u-u_2(\Omega)] \hat{\cal T}(u,\Omega)~,
\label{eq-gslintscri}
\end{equation}
where $\infty$ now refers to a sufficiently late cut of ${\cal I}^+$. Again, this can also be derived directly from the bulk integrated weak-gravity GSL, Eq.~(\ref{eq-weakgslint}), using Eq.~(\ref{eq-wuwu}) to convert from $(w-w_2)$ to $(u-u_2)$. Note that the integral is now over $(u,\Omega)$. Strictly, before taking the limit, $u_2[u,\Omega]$ is the initial cut at the fixed generator $(\rho,\phi)$ determined by $\Omega$ at $u$. This corresponds to a different value of $\Omega$ at $u_2$, but again the angle difference vanishes in the $u_p\to\infty$ limit, by the remarks following Eq.~(\ref{eq-wuwu}).

Since the state above a sufficiently late cut will not differ from the global vacuum state reduced to the same region, the l.h.s.\ of Eq.~(\ref{eq-gslintscri}) is the vacuum-subtracted entropy (Casini entropy) of this semi-infinite region. Hence we can rewrite the equation as
\begin{equation}
\hat S_C[\hat\sigma_2]\leq
\frac{2\pi}{\hbar}\int_{\hat\sigma_2}^{\infty}\! d^2\Omega\, du\, 
[u-u_2(\Omega)]\, \hat{\cal T}(u,\Omega)~.
\label{eq-gslintscrinew}
\end{equation}

\subsection{Quantum Bousso Bound on ${\cal I}^+$}

We now apply the covariant bound on the Casini entropy (the QBB) of a finite portion of $H(u_p)$, Eq.~(\ref{eq-qbb}), to the region defined by cuts $\hat\sigma_1$, $\hat\sigma_2$ on ${\cal I}^+$. Taking the limit as $u_p\to\infty$, recalling that $u$ becomes an affine parameter in the limit, and using Eqs.~(\ref{eq-t}) and (\ref{eq-gg}), we obtain the QBB on ${\cal I}^+$:
\begin{equation}
\hat S_C\leq \frac{2\pi}{\hbar}\Delta \hat K~,
\label{eq-sdkinf}
\end{equation}
where we identify the limit of the modular energy
\begin{equation}
\Delta \hat K =  \int d^2\Omega \int_{u_1(\Omega)}^{u_2(\Omega)} du\, \hat g(u)\, \hat{\cal T}(u,\Omega)
\label{eq-dkinf}
\end{equation}
as the expectation value of a modular Hamiltonian on ${\cal I}^+$. Moreover, $\hat S_C$ is defined as the limit of the vacuum-subtracted entropies on the surfaces $H(u_p)$. It may be identified as the vacuum-subtracted entropy of the asymptotic quantum state on ${\cal I}^+$, restricted to the region between $\sigma_1$ and $\sigma_2$.

The QBB is more subtle than the other bounds due to qualitative differences between the free and interacting cases.\footnote{Another peculiarity of the QBB, in the form of Eq.~(\ref{eq-qbb}), is the requirement that the classical expansion be of semi-definite sign; otherwise $8\pi G\Delta K\leq \Delta A$ need not hold. This can be arranged by $O(G/u_p^2)$ deformations of $\sigma_1$~\cite{BouCas14a}. Here we sidestep the issue by using Eq.~(\ref{eq-sdk}) directly.} For interacting bulk fields (such as gravitons), the $u_p\to\infty$ limit may be discontinuous, since interactions turn off near ${\cal I}^+$. We expect that the free weighting function, Eq.~(\ref{eq-gfree}), will be the one relevant to the asymptotic limit:
\begin{equation}
\hat g(u) = \frac{(u_2-u)(u-u_1)}{u_2-u_1}~,
\end{equation}
even for fields which interact at any finite value of $u_p$. 
These points bear further investigation. It would be interesting to study the algebra of operators on ${\cal I}^+$ directly. However, they do not affect the validity of Eq.~(\ref{eq-sdkinf}) in the general form stated above.

\acknowledgments
It is a pleasure to thank Andy Strominger for discussions and initial collaboration; see Ref.~\cite{KapRac16} for related work by Kapec, Raclariu, and Strominger. I also thank H.~Casini, Z.~Fisher, E.~Flanagan, I.~Halpern, G.~Horowitz, D.~Kapec, J.~Koeller, J.~Maldacena, and R.~Wald for discussions and correspondence. I am grateful to Z.~Fisher for producing Fig.~\ref{fig-conical} and to N.~Engelhardt, E.~Flanagan, A.~Wall, and R.~Wald for helpful comments on a draft and to H.~Casini for correcting a definition. This work was supported in part by the Berkeley Center for Theoretical Physics, by the National Science Foundation (award numbers 1214644, 1521446, and 1316783), by FQXi, and by the US Department of Energy under contract DE-AC02-05CH11231.


\bibliographystyle{utcaps}
\bibliography{all}
\end{document}